\documentclass[a4paper,11pt]{article}
\usepackage{pos}
\usepackage{wrapfig}
\usepackage{bbm}
\usepackage{natbib}
\usepackage{amsmath}
\usepackage[mathscr]{euscript}
\usepackage{graphicx}% Include figure files
\usepackage{dcolumn}% Align table columns on decimal point
\usepackage{bm}% bold math
\usepackage[ruled, vlined]{algorithm2e}
\usepackage{silence}
\usepackage{subcaption}
\usepackage[normalem]{ulem}
\definecolor{purple}{HTML}{571E65}
\definecolor{teal}{HTML}{2C9691}
\definecolor{green}{HTML}{66CB69}
\definecolor{altblue}{HTML}{4F6496}
\definecolor{yellow}{HTML}{FDE726}%{D0BE1E}
\definecolor{darkblue}{HTML}{3B44C0}
\definecolor{darkred}{HTML}{b40626}
\definecolor{blue}{HTML}{007fff}
\definecolor{red}{HTML}{FF2052}

\newcommand{\x}{\mathbf{x}}
\newcommand{\p}{\mathbf{v}}

\newcommand{\green}[1]{\textcolor{green}{{#1}}}
\newcommand{\teal}[1]{\textcolor{teal}{{#1}}}
\newcommand{\purple}[1]{\textcolor{purple}{{#1}}}
\newcommand{\yellow}[1]{\textcolor{yellow}{{#1}}}
\newcommand{\altblue}[1]{\textcolor{altblue}{{#1}}}
\newcommand{\darkblue}[1]{\textcolor{darkblue}{{#1}}}
\newcommand{\darkred}[1]{\textcolor{darkred}{{#1}}}
\newcommand{\mask}{\textcolor{blue}{{m}}}
\newcommand{\maskbar}{\textcolor{red}{\bar{m}}}

\newcommand{\xmask}{\textcolor{blue}{{x}_{m}}}
\newcommand{\xmaskbar}{\textcolor{red}{x_{\bar{m}}}}

\newcommand{\acceptProb}{%
            A(\xi^{\ast}|\xi) \equiv \min\left\{1,
            \frac{p(\xi^{\ast})}{p(\xi)}\right\}
    }

\title{LeapfrogLayers: A Trainable Framework for Effective Topological Sampling}
\ShortTitle{LeapfrogLayers}

\author*[a]{Sam Foreman}
\author[a,b]{Xiao-Yong Jin}
\author[a,b]{James C. Osborn}

\affiliation[a]{Leadership Computing Facility, Argonne National Laboratory,\\
  Lemont, IL, USA}
\affiliation[b]{Computational Science Division, Argonne National Laboratory,\\
  Lemont, IL, USA}

\emailAdd{foremans@anl.gov}
\emailAdd{xjin@anl.gov}
\emailAdd{osborn@alcf.anl.gov}

\abstract{
    We introduce LeapfrogLayers, an invertible neural network architecture that
    can be trained to efficiently sample the topology of a 2D $U(1)$ lattice
    gauge theory.
    We show an improvement in the integrated autocorrelation time of the
    topological charge when compared with traditional HMC, and look at how
    different quantities transform under our model.
    Our implementation is open source, and is publicly available on GitHub at
    \href{https://www.github.com/saforem2/l2hmc-qcd}{github.com/saforem2/l2hmc-qcd}.
}

\FullConference{
  The 38th International Symposium on Lattice Field Theory\\
  26-30 July 2021\\
  Zoom / Gather @ MIT, Cambridge MA, USA\\
}

\begin{document}
\maketitle

\section{\label{sec:intro}Introduction}
The main task in lattice field theory calculations is to evaluate
integrals of the form
\begin{equation}
  \langle \mathcal{O} \rangle =
  \int \left[ \mathcal{D} \x \right] \mathcal{O}(\x) p(\x),
    \label{eq:density_integral}
\end{equation}
for some normalized multivariate target distribution \(p(\x) = e^{-S(\x)}/Z\).
We can approximate the integral using Markov Chain Monte Carlo (MCMC) sampling
techniques.
This is done by sequentially generating a chain of configurations \(\{\x_{1},
\x_{2}, \ldots \x_{N}\}\), with \(\x_{i}\) occurring according to the stationary distribution \(p(\x)\), and averaging the value of
the function \(\mathcal{O}(\x)\) over the chain
\begin{equation}
  \overline{\mathcal{O}} = \frac{1}{N} \sum_{n=1}^{N} \mathcal{O}(\x_n)  .
\end{equation}
Accounting for correlations between states in the chain, the sampling variance
%of this estimator
is given by
\begin{equation}
  \sigmaup^{2} = \frac{\tau_{\mathcal{O}}^{\mathrm{int}}}{N}\sum_{n=1}^{N}
  \left[ \mathcal{O}(\x_n) - \overline{\mathcal{O}} \right]^2
\end{equation}
where \(\tau^{\mathrm{int}}_{\mathcal{O}}\) is the integrated autocorrelation
time.
This quantity can be interpreted as the additional time required for
these induced correlations to become negligible.

While our main target for improved gauge field generation is full QCD,
here we restrict our work to the simpler 2D U(1) pure-gauge lattice
theory.
There are already many efficient alternatives for simulating this particular theory
(see e.g.~\cite{Eichhorn:2021ccz} and references therein for a comparison),
however scaling up to full QCD remains challenging.
As the preferred method for full QCD,
here HMC is our baseline for performance comparison.
The method presented here can be extended to full QCD in a relatively
straightforward manner, though the cost of training and efficiency are
still open questions.
\subsection{\label{subsec:qfreezing}Charge Freezing}
The ability to efficiently generate independent configurations is currently
a major bottleneck for lattice simulations.
The 2D U(1) gauge theory considered here has sectors of topological charge that
become separated by large potential barriers for increasing $\beta$,
similar to full QCD.

\begin{wrapfigure}[21]{r}{0.46\textwidth}
  %\vspace{0.5\baselineskip}
  \centering
    \includegraphics[width=0.45\textwidth]{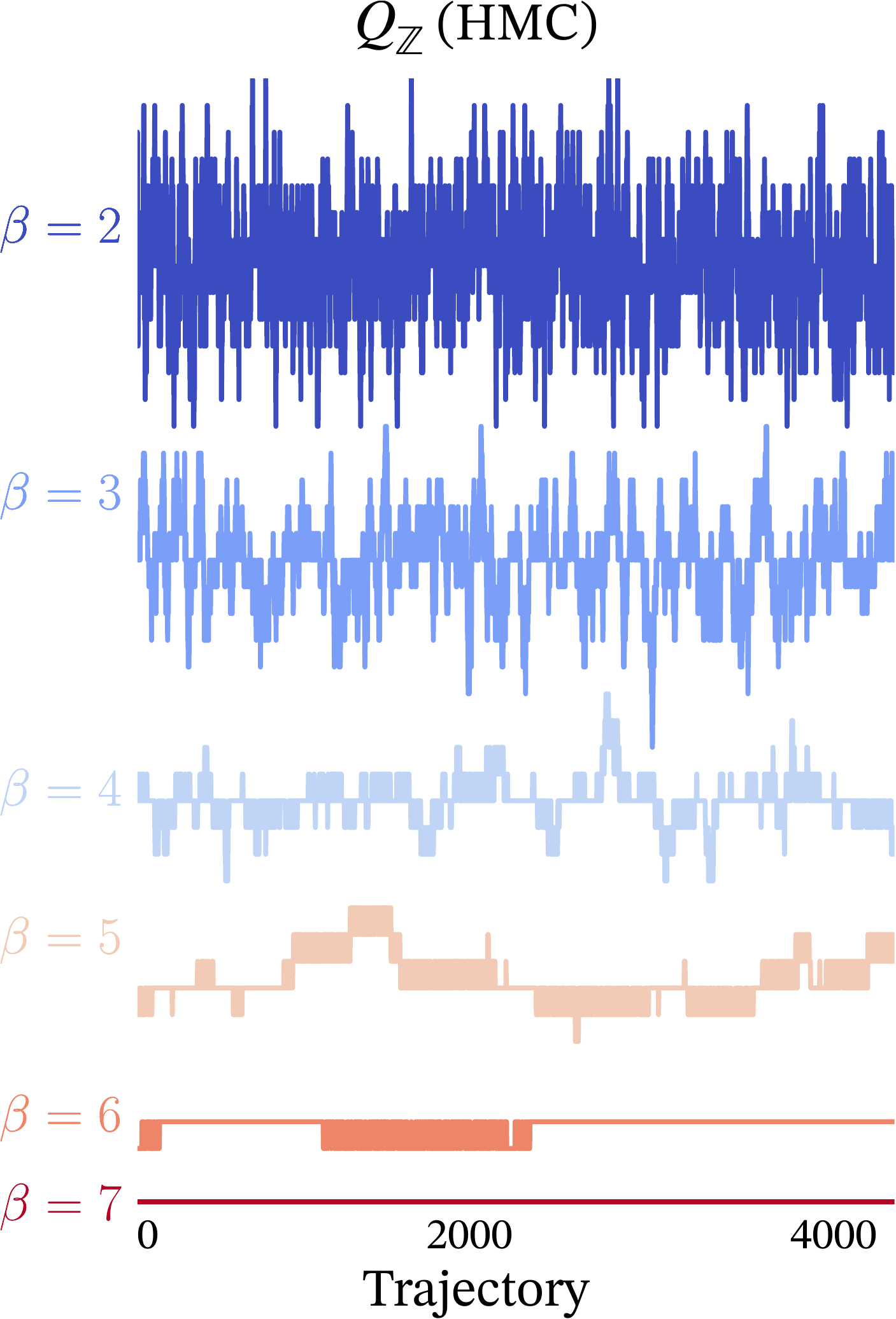}
  \caption{\label{fig:qfreezing}Illustration of the topological charge,
    \(Q_{\mathbb{Z}}\), freezing as \(\beta : \darkblue{2} \rightarrow \darkred{7}\)
  for traditional HMC.}%
\end{wrapfigure}

The theory is defined in terms of the link variables
\(U_{\mu}(x) = e^{i x} \in U(1)\) with \(x \in [-\pi, \pi)\).
Our target distribution is given by \(p(\x)\propto e^{-S_{\beta}(\x)}\), where
\(S_{\beta}(\x)\) is the Wilson action
\begin{equation}
  S_{\beta}(\x) = \beta \sum_{P} 1 - \cos{x_{P}},
\end{equation}
defined in terms of the sum of the gauge variables around the elementary
plaquette, \(x_{P} \equiv x_{\mu}(n) + x_{\nu}(n + \hat{\mu}) -
x_{\mu}(n+\hat{\nu}) - x_{\nu}(n)\), starting at site \(n\) of the
lattice, and \(\sum_{P}\) denotes the sum over all such plaquettes.
For a given lattice configuration, we can calculate the topological charge \(Q
\in \mathbb{Z}\) using
\begin{equation}
    Q_{\mathbb{Z}} = \frac{1}{2\pi}\sum_{P}\left\lfloor x_{P} \right\rfloor,\text{ where }
    \left\lfloor x_{P} \right\rfloor \equiv x_{P} - 2\pi
    \left\lfloor\frac{x_{P}+\pi}{2\pi}\right\rfloor.
\end{equation}
As \(\beta \rightarrow \infty\),
we see that the value of \(Q_\mathbb{Z}\) remains fixed for large
durations of the simulation.
This freezing can be quantitatively measured by introducing the \emph{tunneling
rate}
(as measured between subsequent states in our chain \(i\) and
\(i+1\) )
\begin{equation}
  \overline{\delta Q}_{\mathbb{Z}} ~,~\mathrm{with}~~
  \delta Q_{\mathbb{Z},i} \equiv \left|Q_{\mathbb{Z},i+1} - Q_{\mathbb{Z},i}\right|
    \in \mathbb{Z},
\end{equation}
which serves as a measure for how efficiently our chain is able to jump
(tunnel) between sectors of distinct topological charge.
From Figure~\ref{fig:qfreezing}, we can see that
\(\overline{\delta Q}_{\mathbb{Z}} \rightarrow 0\) as
\(\beta\rightarrow \infty\).
\vspace{-\baselineskip}
\section{\label{sec:hmc}Hamiltonian Monte Carlo (HMC)}
The Hamiltonian Monte Carlo algorithm begins by introducing a fictitious
momentum, \(v\), for each coordinate variable, $x$, typically taken
from an independent normal distribution.
This allows us to write the joint target density of the \(\xi \equiv (\x, \p)\)
system as
\begin{equation}
    p(\x, \p) = p(\x) p(\p) = e^{-S_{\beta}(\x)} e^{-\p^2 / 2} = e^{-\mathcal{H} (\x, \p)}
\end{equation}
where \(\mathcal{H}(\xi) = \mathcal{H}(\x, \p) = S_{\beta}(\x) + \frac{1}{2} \p^2
\) is the Hamiltonian of the system.
We use the \emph{leapfrog integrator} to approximately numerically integrate
Hamilton's equations
\begin{equation}
    \dot{\x} = \frac{\partial \mathcal{H}}{\partial \p}\quad \text{,} \quad
    \dot{\p} = - \frac{\partial \mathcal{H}}{\partial \x}
\end{equation}
along iso-probability contours of \(\mathcal{H} =\text{const.}\) from \(\xi =
(\x,\p)\rightarrow (\x^{\ast}, \p^{\ast}) = \xi^{\ast}\).
The error in this integration is then corrected by a Metropolis-Hastings (MH)
accept/reject step.
\subsection{\label{subsec:lfint}HMC algorithm with Leapfrog Integrator}
\begin{figure}[htpb]
    \centering
    \includegraphics[width=\linewidth]{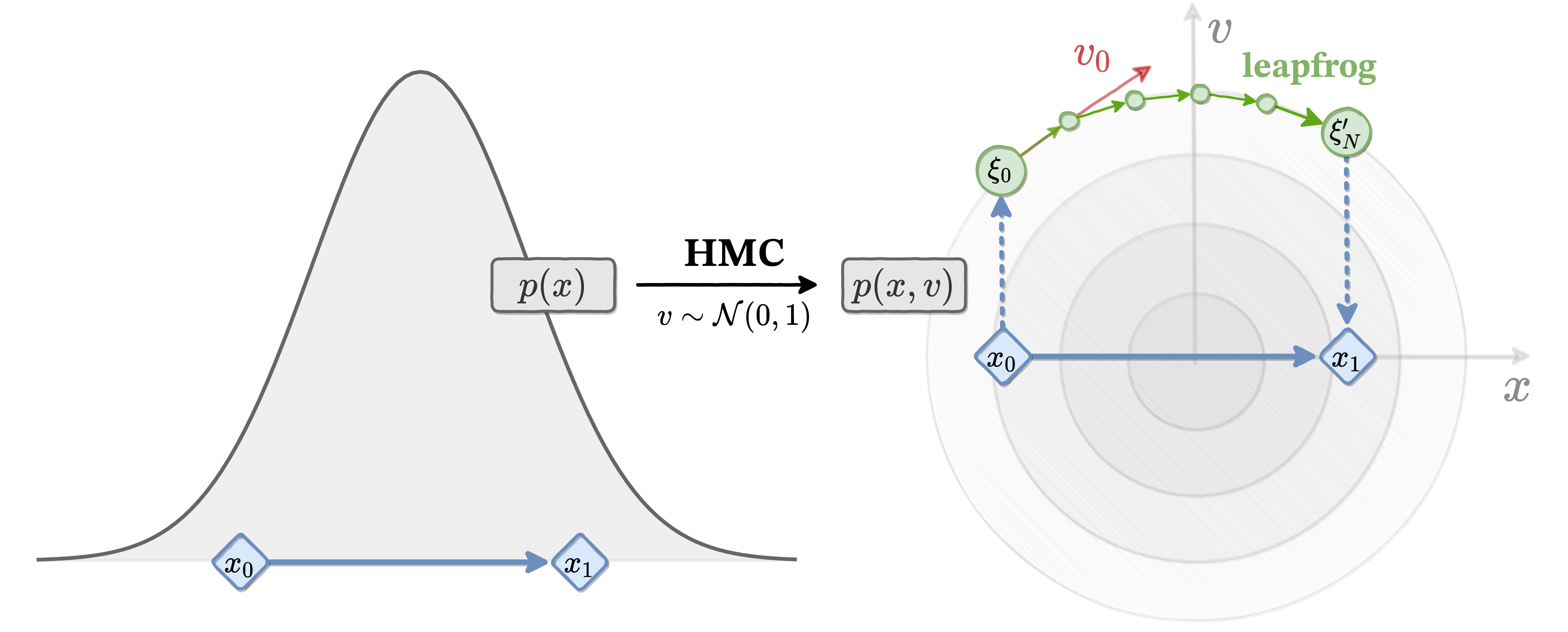}
    \caption{\label{fig:hmc}High-level overview of the HMC algorithm.}
\end{figure}

\begin{enumerate}
    \item Starting from \(\x_{i}\), resample the momentum \(\p_{i}\sim
        \mathcal{N} (0, \mathbbm{1})\) and construct the state \(\xi =
        (\x_i, \p_i)\).
    \item Generate a \emph{proposal configuration} \(\xi^{\ast}\) by
        integrating \(\dot\xi\) along \(\mathcal{H} = \mathrm{const.}\)
        for \(N\) leapfrog steps.
        i.e.
        \begin{equation}
            \xi \rightarrow \xi_{1}\rightarrow\ldots\rightarrow
            \xi_{N} \equiv \xi^{\ast},
        \end{equation}
        where a single leapfrog step \(\xi_{i} \rightarrow \xi_{i+1}\) above
        consists of:
        \begin{equation}
            \textbf{ (a.) }%
              \p'\leftarrow \p - \frac{\varepsilon}{2}\partial_{\x} S(\x),
            \quad\quad
            \textbf{ (b.) }%
              \x' \leftarrow \x + \varepsilon \p',
            \quad\quad
            \textbf{ (c.) }%
              \p'' \leftarrow \p' - \frac{\varepsilon}{2}\partial_{\x} S(\x).
        \end{equation}
    \item At the end of the trajectory, accept or reject the proposal
        configuration \(\xi^{\ast}\) using the MH test
        \begin{equation}
            \x_{i+1} \leftarrow
            \begin{cases}
                \x^{\ast}\text{ with probability } \acceptProb \\
                \x_{i}\text{ with probability } 1 - A(\xi^{\ast}|\xi).
            \end{cases}
        \end{equation}
\end{enumerate}
An illustration of this procedure can be seen in Figure~\ref{fig:hmc}.
\subsection{\label{subsec:hmc_issues}Issues with HMC}
The HMC sampler is known to have difficulty traversing areas
where the action is large.
This makes it less likely for the system to go into low probability regions,
and effectively bounds the system within potential barriers,
resulting in poor performance for distributions which have
multiple isolated modes.
This is particularly relevant in the case of sampling topological quantities in
lattice gauge models.
\section{\label{sec:l2hmc}Generalizing HMC: LeapfrogLayers}
In order to preserve the exact stationary distribution of the Markov Chain, our update must be
explicitly reversible with a tractable Jacobian determinant.
To simplify notation, we introduce two functions, \(\Gamma\) \((\Lambda)\) to
denote the \(\p\) \((\x)\) updates.
As in HMC, we follow the general pattern of performing alternating updates of
\(\p\) and \(\x\).

We can ensure
the Jacobian is simple
by splitting the \(\x\) update into two
parts and sequentially updating complementary subsets using a binary mask
\(\mask\) and its complement \(\maskbar\).
As in~\cite{2017arXiv171109268L}, we introduce \(d \sim \mathcal{U} (+, -)\),
distributed independently of both \(\x\) and \(\p\), to determine the
``direction'' of our update\footnote{%
  As indicated by the superscript \(\pm\) on \(\Gamma^{\pm}, \Lambda^{\pm}\) in
  the update functions.
}.
Here, we associate \(+\) \((-)\) with the forward (backward) direction and note
that running sequential updates in opposite directions has the effect of
inverting the update.
We denote the complete state by \(\xi = (\x, \p, d)\), with target density
given by \(p(\xi) = p(\x) p(\p) p(d)\).

Explicitly, we can write this series of updates as\footnote{%
  Here we denote by \(\xmask = \mask \odot \x\) and 
  \(\xmaskbar = \maskbar \odot \x\) with \(\mathbbm{1} = \mask + \maskbar\).
}
\begin{align}
    \nonumber
    \textbf{(a.)}\,\,\, \p'
        \gets \Gamma^{\pm}\left[\p;\, \zeta_{\p} \right]\hspace{80pt}
    &\textbf{(b.)}\,\,\, \x'
        \gets \mask \odot \x + \maskbar \odot \Lambda^{\pm} [\xmaskbar; \zeta_{\bar{\x}}] \\
    \nonumber
    \textbf{(c.)}\,\, \x''
        \gets \maskbar \odot \x' + \mask \odot \Lambda^{\pm} [\xmask; \zeta_{\x'}] \quad
    &\textbf{(d.)}\,\, \p''
        \gets \Gamma^{\pm} [\p', \zeta_{\p'}]
\end{align}
where \(\zeta_{\bar{\x}} = [\maskbar\odot \x, \p]\), \(\zeta_{\x} = [\mask \odot \x,
\p]\) \((\zeta_{\p} = [\x, \partial_{\x} S(\x)])\) is independent of \(\x\) \((\p)\)
and is passed as input to the update functions \(\Lambda^{\pm}\)
\((\Gamma^{\pm})\).
The acceptance probability
\begin{equation}
A(\xi^{\ast} | \xi) =
\min\left\{1, \frac{p(\xi^{\ast})}{p(\xi)} \mathcal{J}(\xi^{\ast},\xi) \right\}
\end{equation}
now includes a Jacobian factor $\mathcal{J}(\xi^{\ast},\xi)$
which allows the inclusion of non-symplectic update steps.
The Jacobian comes from the $v$ ($x$) scaling term in the $v$ ($x$) update, and
is easily calculated.
\begin{figure}[htpb]
    \centering
    \begin{subfigure}[t]{0.40\linewidth}
        \includegraphics[width=\linewidth]{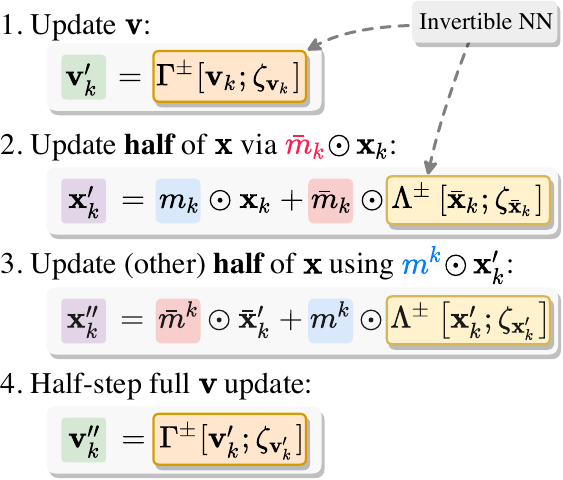}
        \caption{\label{subfig:updates}Generalized leapfrog update.}
    \end{subfigure}
    \hfill
    \begin{subfigure}[t]{0.55\linewidth}
        \includegraphics[width=\linewidth]{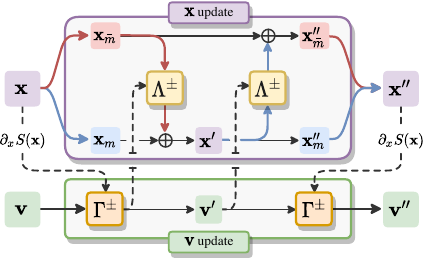}
        \caption{\label{subfig:lfupdate}Illustration of the data flow
        through a leapfrog layer.}
    \end{subfigure}
    % \vspace{1em}
    \begin{subfigure}[t]{\linewidth}
        \includegraphics[width=\linewidth]{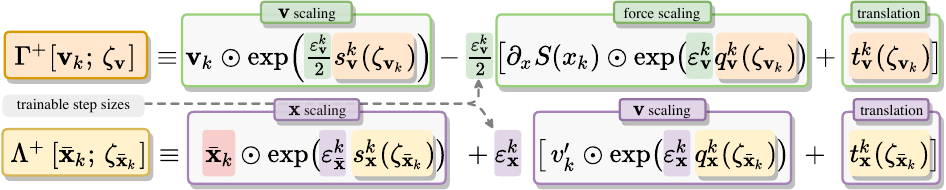}
        \caption{\label{subfig:network_fns}Detailed view of the update
        functions \(\Gamma^{+}, \Lambda^{+}\) for the \(k^{\mathrm{th}}\)
    leapfrog step.}
    \end{subfigure}
    \caption{\label{fig:networks}Illustrations of the generalized
        leapfrog update. In Figure~\ref{subfig:updates},
        ~\ref{subfig:network_fns}, we denote \(\bar{\mathbf{x}}_{k} =
    \maskbar\odot \mathbf{x}_{k}\).}
\end{figure}
\subsection{\label{subsec:networks}Network Details}
Normalizing flows~\cite{2016arXiv160508803D} are an obvious choice for the
structure of the update functions.
These architectures are easily invertible while maintaining a tractable
Jacobian determinant, and have also been shown to be effective at approximating
complicated target distributions in high-dimensional spaces ~\cite{%
    2016arXiv160508803D,% dinhDensityEstimationUsing2016,%
    2017arXiv171109268L,% levyGeneralizingHamiltonianMonte2018,%
    liNeuralNetworkRenormalization2018a,%
    Foreman2018,%
    Kanwar:2020xzo,%
    wehenkelYouSayNormalizing2020,%
    Nicoli:2020njz,%
    Boyda:2020hsi,%
    Li2021ANN,%
    Albergo:2021vyo,%
    Foreman:2021ljl,%
    Foreman:2021ixr%
    %Neklyudov2020OrbitalM,%
    %neklyudovInvolutiveMcmcUnifying2020,%
}.

We maintain separate networks \(\Gamma\), \(\Lambda\) with identical
architectures for updating \(\p\) and \(\x\), respectively.
Without loss of generality, we describe below the details of the \(\x\) update
for the forward \((d = +)\) direction, \(\Lambda^{+}[\xmaskbar;
\zeta_{\bar{\x}}]\)\footnote{
    To get the update for the \(d=-\) direction, we invert the
    update functions and run them in the opposite order.
}.
For simplicity, we describe the data flow through a single leapfrog layer,
which takes as input \(\zeta_{\bar{\x}} = (\xmaskbar, \p)\).
For the 2D \(U(1)\) model, the gauge links are encoded as \([\cos(x),
\sin(x)]\) for \(x \in [-\pi, \pi]\).
Explicitly\footnote{
Here \(\sigma(\cdot)\) is a generic nonlinear activation function
acting independently on the elements of the output vector.
},
\begin{align}
    h_{1} &= \sigma\left(w_{x} \x + w_{v} \p + b_{1}\right)
        \quad \in \mathbb{R}^{n_{1}} \\
    h_{2} &= \sigma\left(w_{2} h_{1} + b_{2}\right)
        \hspace{0.085\textwidth} \in \mathbb{R}^{n_{2}} \\
    \nonumber \vdots & \\
    h_{k} &= \sigma\left(w_{k} h_{k-1} + b_{k-1}\right)
        \hspace{0.04\textwidth} \in \mathbb{R}^{n_{k}} \Longrightarrow \\
    \nonumber
    \textbf{  (a.)  } s_{x} = \lambda_{s} \tanh \left( w_{s} h_{k} + b_{s} \right);
    &\quad\textbf{  (b.)  } t_{x} = w_{t} h_{k} + b_{t};
    \quad\textbf{  (c.)  } q_{x} = \lambda_{q}\tanh\left(w_{q} h_{k} + b_{q}\right);
\end{align}
where the outputs \(s_{x}, t_{x}, q_{x}\) are of the same dimensionality as
\(\x\), and \(\lambda_{s}, \lambda_{q}\) are trainable parameters.
These outputs are then used to update \(\x\), as shown in
Figure~\ref{fig:networks}.

\subsection{\label{subsec:training}Training Details}
Our goal in training the network is to find a sampler that efficiently
jumps between different topological charge sectors.
This can be done by constructing a loss function that maximizes the expected
squared charge difference between the initial \((\xi)\) and proposal
\((\xi^{\ast})\) configuration generated by the sampler.
Explicitly, we define
\begin{equation}
    \mathcal{L}_{\theta}(\xi^{\ast}, \xi)
    = A(\xi^{\ast}|\xi) (Q^{\ast}_{\mathbb{R}} - Q_{\mathbb{R}})^{2}
\end{equation}
where \(Q_{\mathbb{R}} = \frac{1}{2\pi}\sum_{P} \sin x_{P} \in \mathbb{R}\) is
a real-valued approximation of the usual (integer-valued) topological charge
\(Q_{\mathbb{Z}} \in \mathbb{Z}\).
This ensures that our loss function is continuous which helps to simplify the
training procedure.
For completeness, the details of an individual training step are summarized in
Sec~\ref{subsubsec:trainstep}.
\subsubsection{\label{subsubsec:trainstep}Training Step}
\begin{enumerate}
    \item Resample \(\p \sim \mathcal{N} (0, \mathbbm{1})\),
        \(d \sim \mathcal{U} (+, -)\), and construct initial state
        \(\xi = (\x, \p, \pm)\)
    \item Generate the proposal configuration \(\xi^{\ast}\) by passing the
        initial state sequentially through \(N_{\mathrm{LF}}\) \emph{leapfrog
        layers}: \(\xi \rightarrow \xi_{1}
        \rightarrow \ldots \rightarrow \xi_{N_{\mathrm{LF}}} = \xi^{\ast}\)
    \item Compute the Metropolis-Hastings acceptance \(A(\xi^{\ast} | \xi) =
      \min\left\{1, \frac{p(\xi^{\ast})}{p(\xi)}
      \mathcal{J}(\xi^{\ast},\xi) \right\}\)
    \item Evaluate the loss function \(\mathcal{L} \leftarrow
        \mathcal{L}_{\theta}(\xi^{\ast}, \xi)\), and back propagate gradients to
        update weights
    \item Evaluate Metropolis-Hastings criteria and assign the next state in
        the chain according to
        \(\x_{t+1} \leftarrow \begin{cases}%
            \x^{\ast} \text{ with prob. } A(\xi^{\ast}|\xi) \\
            \x \text{ with prob. } 1 - A(\xi^{\ast}|\xi).
        \end{cases}\)
\end{enumerate}
\subsection{\label{subsec:annealing}Annealing}
As an additional tool to help improve the quality of the trained sampler, we
scale the action during the $N_{\mathrm{T}}$ training steps using the target
distribution \(p_{t}(x) \propto e^{-\gamma_{t} S(x)}\) for \(t = 0, 1, \ldots,
N_{\mathrm{T}}\).
The scale factors, $\gamma_{t}$, monotonically increase according to an
\emph{annealing schedule} up to $\gamma_{N_{\mathrm{T}}} \equiv 1$, with
\(|\gamma_{i+1} - \gamma_{i}| \ll 1\).
For \(\gamma_i < 1\), this helps to rescale (shrink) the energy barriers
between isolated modes, allowing the training to experience sufficient
tunneling even when the final distribution is difficult to sample.
\clearpage
\section{\label{sec:results}Results}
\begin{wrapfigure}[15]{r}{0.58\textwidth}
    \vspace{-\baselineskip}
    \centering
    \includegraphics[width=0.58\textwidth]{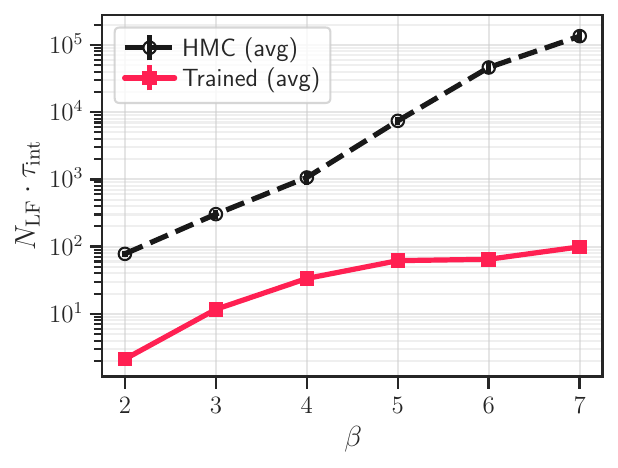}
    \caption{\label{fig:autocorr}Estimated
    \(\tau_{\mathrm{int}}^{\mathcal{Q}_{\mathbb{R}}}\) vs \(\beta\).}
\end{wrapfigure}
\begin{figure}[htpb]
    \centering
    \begin{subfigure}[t]{0.325\textwidth}
        \includegraphics[width=\textwidth]{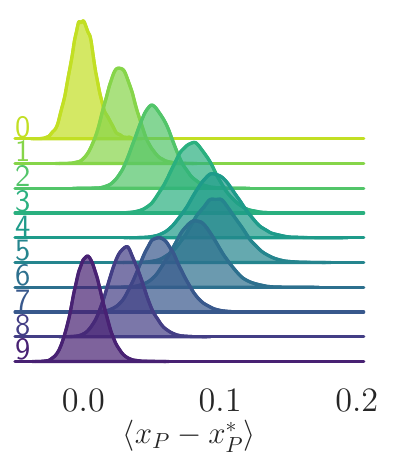}
        \caption{\label{fig:plaqsf_ridgeplot}Deviation in \(x_{P}\).}
    \end{subfigure}
    \hfill
    \begin{subfigure}[t]{0.32\textwidth}
        \includegraphics[width=\textwidth]{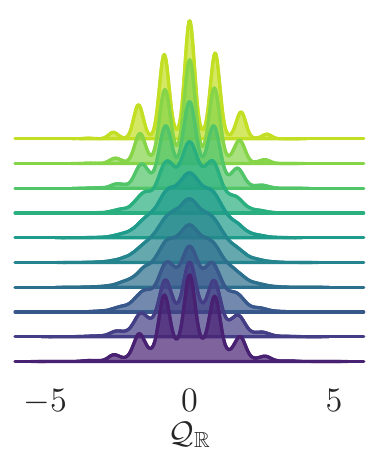}
        \caption{\label{fig:sinQf_ridgeplot}Topological charge mixing
        \(Q_{\mathbb{R}}\).}
    \end{subfigure}
    \hfill
    \begin{subfigure}[t]{0.318\textwidth}
        \includegraphics[width=\textwidth]{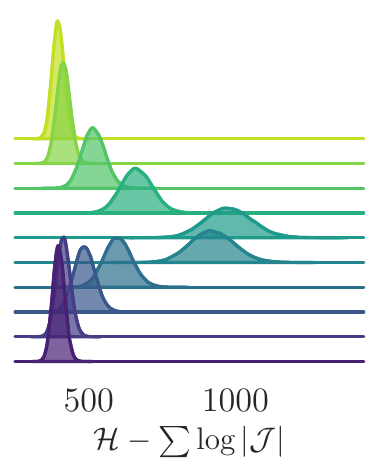}
        \caption{\label{fig:hwf_ridgeplot}Artificial influx of energy.}
    \end{subfigure}
    \caption{\label{fig:ridgeplots}Evolution of different quantities over a
    single trajectory consisting of $N_{\mathrm{LF}} = 10$ leapfrog steps.}%
\end{figure}
\begin{figure}[htpb]
    \begin{subfigure}[b]{0.45\textwidth}
        \includegraphics[width=\linewidth]{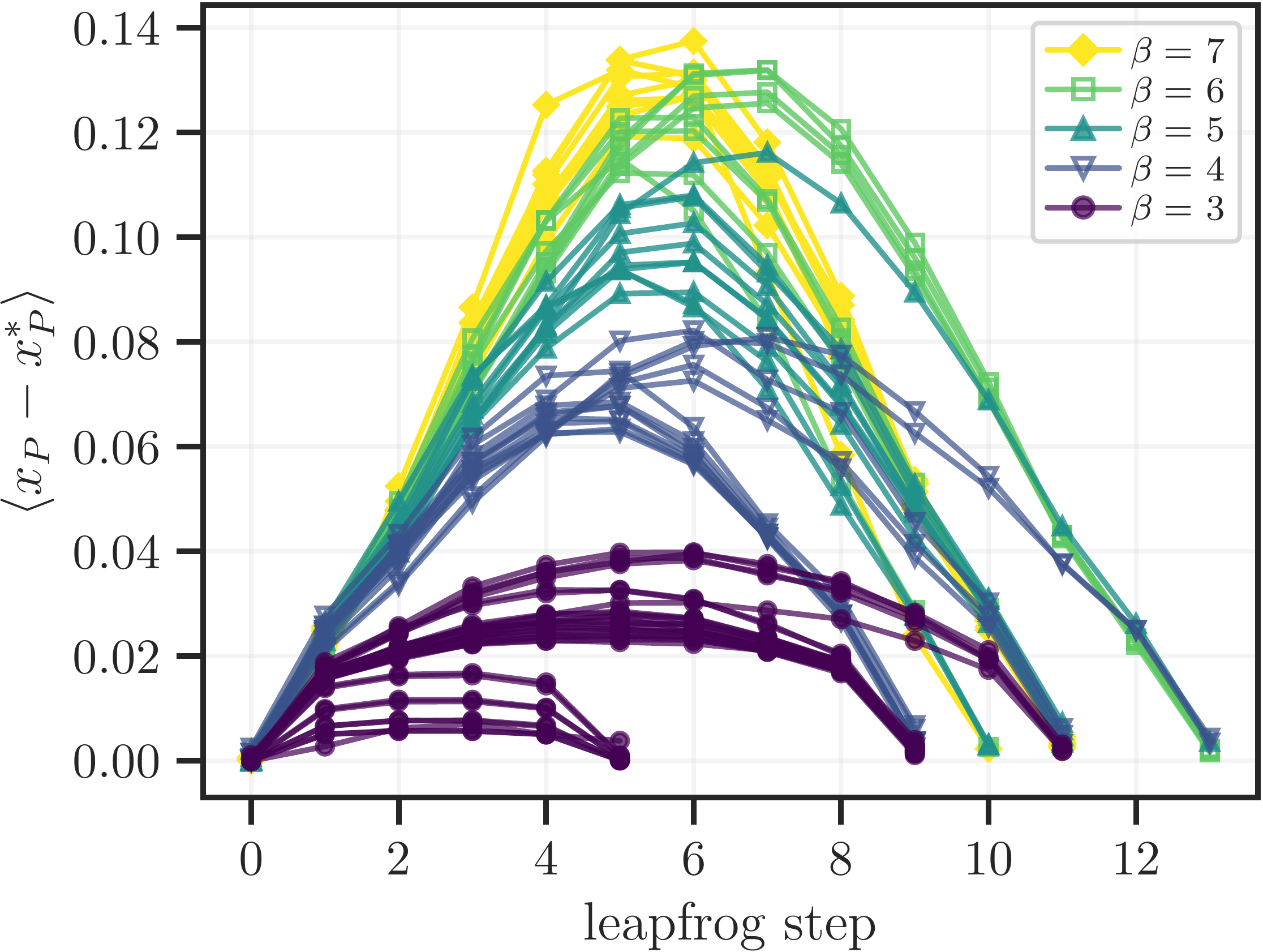}
        \caption{\label{subfig:dplaqsf}The deviation of \(x_{P}\) from the
        \(V\rightarrow\infty\) limit, \(x_{P}^{\ast}\).}
    \end{subfigure}
    \hfill
    \begin{subfigure}[b]{0.45\textwidth}
        \includegraphics[width=\linewidth]{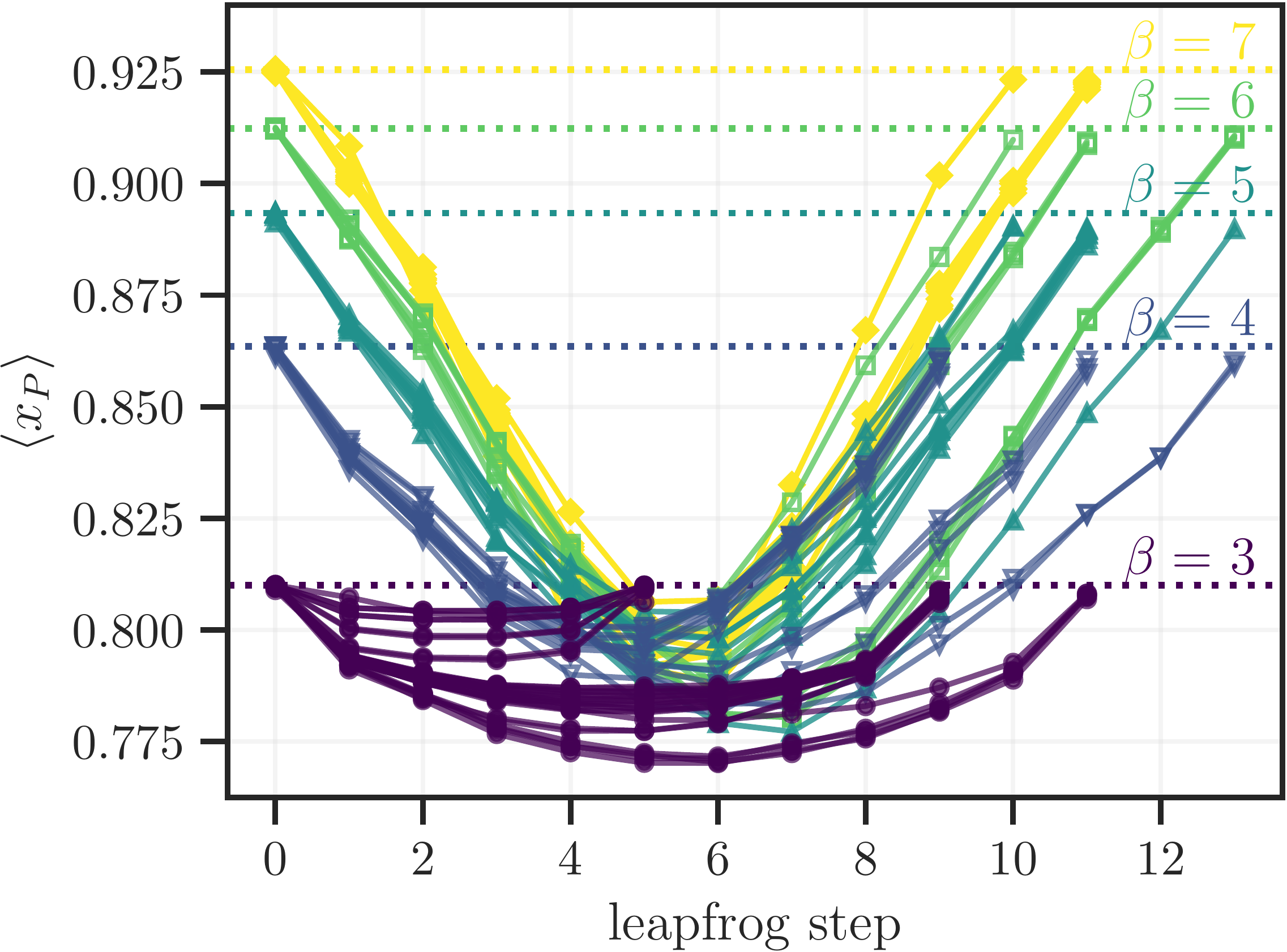}
        \caption{\label{subfig:plaqsf}Note \(x^{\ast}_{P}\) is indicated by the
        horizontal dashed lines.}
    \end{subfigure}
    \caption{\label{fig:plaqsf}Plots showing how the plaquette varies
        over a single trajectory for models trained at \(\beta =
    \purple{3}, \altblue{4}, \teal{5}, \green{6}, \yellow{7}\).}
\end{figure}

We can measure the performance of our approach by comparing the integrated
autocorrelation time against traditional HMC.
We see in Figure~\ref{fig:autocorr} that our estimate of the integrated
autocorrelation time is much shorter for the trained model across \(\beta \in
[2, 7]\).
To better understand how these transformations effect the HMC update, we can
look at how various quantities change over the course of a trajectory, as shown
in Figure~\ref{fig:ridgeplots}.
We see from the plot of \(\mathcal{H} - \sum \log | \mathcal{J} |\) in
Figure~\ref{fig:ridgeplots}c that the trained sampler artificially
increases the energy towards the middle of the trajectory.
This is analogous to reducing the coupling \(\beta\) during the trajectory, as
can be seen in Figure~\ref{subfig:plaqsf}.
In particular, we can see that for \(\beta = 7\), the value of the plaquette at
in the middle of the trajectory roughly corresponds to the expected value at
\(\beta = 3\), indicated by the horizontal dashed line.
This effective reduction in the coupling constant allows our sampler to mix
topological charge values in the middle of the trajectory, before returning to
the stationary distribution at the end, as can be seen in
Figure~\ref{fig:sinQf_ridgeplot}.
By looking at the variation in the average plaquette \(\langle x_{P} -
x^{\ast}_{P}\rangle\) over a trajectory for models trained at multiple values
of beta we are able to better understand how our sampler behaves.
This allows our trajectories to explore new regions of the target distribution
which may have been previously inaccessible.
\begin{figure}[htpb]
    \includegraphics[width=\textwidth]{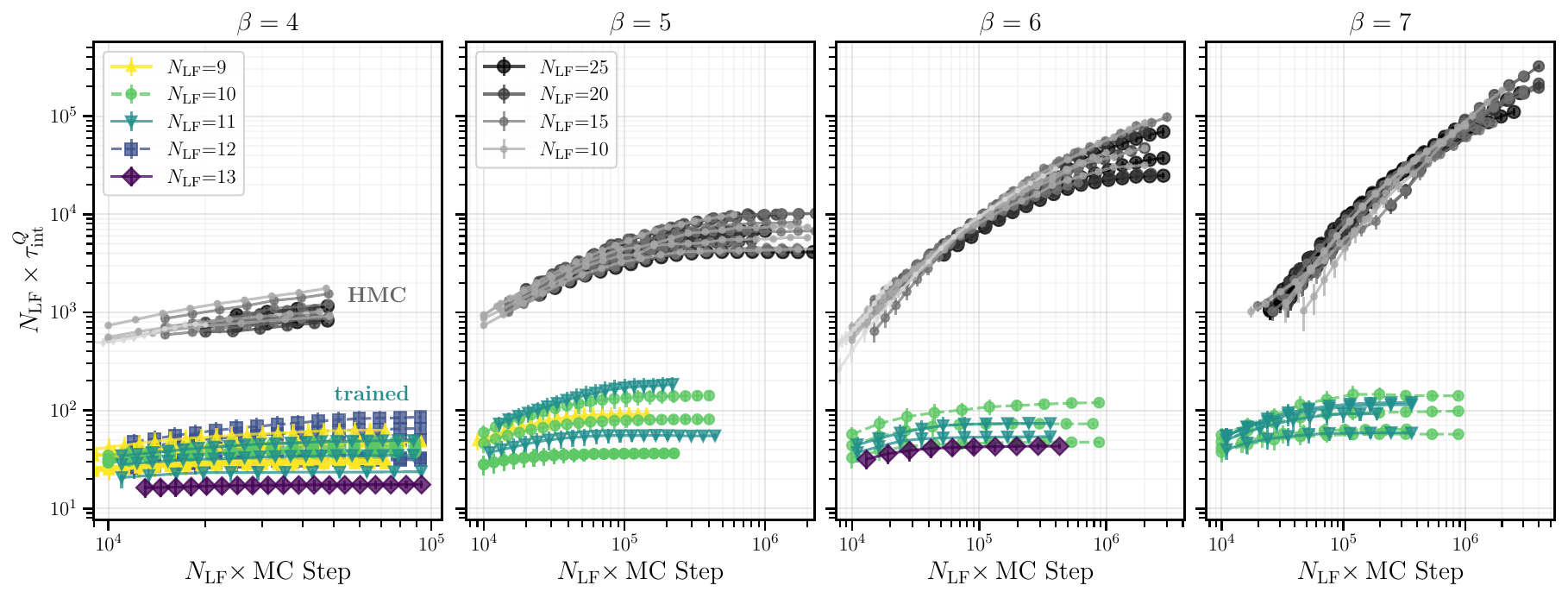}
    \caption{\label{fig:tint}Comparison of the integrated autocorrelation time
        for trained models vs HMC with different trajectory lengths,
    \(N_{\mathrm{LF}}\), at \(\beta = 4, 5, 6, 7\) (left to right).}
\end{figure}

\section{\label{sec:conclusion}Conclusion}
In this work we have proposed a generalized sampling procedure for HMC that can
be used for generating gauge configurations in the 2D \(U(1)\) lattice model.
We showed that our trained model is capable of outperforming traditional
sampling techniques across a range of inverse coupling constants, as measured
by the integrated autocorrelation time of the topological charge.
By looking at the evolution of different quantities over the generalized
leapfrog trajectory, we are able to gain insight into the mechanism driving
this improved behavior.
\section{\label{sec:ack}Acknowledgments}
This research was supported by the Exascale Computing Project (17-SC-20-SC), a
collaborative effort of the U.S. Department of Energy Office of Science and the
National Nuclear Security Administration.
This research was performed using resources of the Argonne Leadership Computing
Facility (ALCF), which is a DOE Office of Science User Facility supported under
Contract DE\_AC02--06CH11357. 
This work describes objective technical results and analysis.
Any subjective views or opinions that might be expressed in the work do not
necessarily represent the views of the U.S. DOE or the United States
Government.
Results presented in this research were obtained using the Python
\citep{van1995python}, programming language and its many data science libraries
\cite{%
    matplotlib,%
    harris2020array,%
    tensorflow2015-whitepaper,%
    ipython4160251%
}.
%
% Explicitly, if we denote by \(M^{\pm} = \left(\begin{smallmatrix}\Gamma^{\pm} &
% \cdot \\ \cdot & \Lambda^{\pm}\end{smallmatrix}\right)\), we have that 
% %
% \begin{equation}
%     \xi^{\pm} = M^{\pm}\xi \Longrightarrow \xi = M^{\mp}\xi^{\pm}.
% \end{equation}
% %

\bibliographystyle{JHEP}
\bibliography{main,ZoteroLibrary}

\providecommand{\href}[2]{#2}\begingroup\raggedright\begin{thebibliography}{10}

\bibitem{Eichhorn:2021ccz}
T.~Eichhorn and C.~Hoelbling, \emph{Comparison of topology changing update
  algorithms},  \href{https://arxiv.org/abs/2112.05188}{{\ttfamily
  2112.05188}}.

\bibitem{2017arXiv171109268L}
D.~{Levy}, M.D.~{Hoffman} and J.~{Sohl-Dickstein}, \emph{{Generalizing
  Hamiltonian Monte Carlo with Neural Networks}}, {\emph{arXiv e-prints} (2017)
  arXiv:1711.09268} [\href{https://arxiv.org/abs/1711.09268}{{\ttfamily
  1711.09268}}].

\bibitem{2016arXiv160508803D}
L.~{Dinh}, J.~{Sohl-Dickstein} and S.~{Bengio}, \emph{{Density estimation using
  Real NVP}}, {\emph{arXiv e-prints} (2016) arXiv:1605.08803}
  [\href{https://arxiv.org/abs/1605.08803}{{\ttfamily 1605.08803}}].

\bibitem{liNeuralNetworkRenormalization2018a}
S.-H.~Li and L.~Wang, \emph{Neural network renormalization group},
  \href{https://doi.org/10.1103/physrevlett.121.260601}{\emph{Physical Review
  Letters} {\bfseries 121} (2018) }.

\bibitem{Foreman2018}
S.~Foreman, \emph{Learning Better Physics: A Machine Learning Approach to
  Lattice Gauge Theory}, Ph.D. thesis, The University of Iowa, 2019.
\newblock 10.17077/etd.500b-30qp.

\bibitem{Kanwar:2020xzo}
G.~Kanwar, M.S.~Albergo, D.~Boyda, K.~Cranmer, D.C.~Hackett, S.~Racani\`ere
  et~al., \emph{{Equivariant flow-based sampling for lattice gauge theory}},
  \href{https://doi.org/10.1103/PhysRevLett.125.121601}{\emph{Phys. Rev. Lett.}
  {\bfseries 125} (2020) 121601}
  [\href{https://arxiv.org/abs/2003.06413}{{\ttfamily 2003.06413}}].

\bibitem{wehenkelYouSayNormalizing2020}
J.~Dumas, A.~Wehenkel, D.~Lanaspeze, B.~Cornélusse and A.~Sutera, \emph{You
  say normalizing flows i see bayesian networks},
  \href{https://arxiv.org/abs/2006.00866}{{\ttfamily 2006.00866}}.

\bibitem{Nicoli:2020njz}
K.A.~Nicoli, C.J.~Anders, L.~Funcke, T.~Hartung, K.~Jansen, P.~Kessel et~al.,
  \emph{Estimation of {{Thermodynamic Observables}} in {{Lattice Field
  Theories}} with {{Deep Generative Models}}},
  \href{https://arxiv.org/abs/2007.07115}{{\ttfamily 2007.07115}}.

\bibitem{Boyda:2020hsi}
D.~Boyda, G.~Kanwar, S.~Racani\`ere, D.J.~Rezende, M.S.~Albergo, K.~Cranmer
  et~al., \emph{{Sampling using $SU(N)$ gauge equivariant flows}},
  \href{https://doi.org/10.1103/PhysRevD.103.074504}{\emph{Phys. Rev. D}
  {\bfseries 103} (2021) 074504}
  [\href{https://arxiv.org/abs/2008.05456}{{\ttfamily 2008.05456}}].

\bibitem{Li2021ANN}
Z.~Li, Y.~Chen and F.T.~Sommer, \emph{A neural network {{MCMC}} sampler that
  maximizes proposal entropy},
  \href{https://arxiv.org/abs/2010.03587}{{\ttfamily 2010.03587}}.

\bibitem{Albergo:2021vyo}
M.S.~Albergo, D.~Boyda, D.C.~Hackett, G.~Kanwar, K.~Cranmer, S.~Racani\`{e}re
  et~al., \emph{Introduction to normalizing flows for lattice field theory},
  \href{https://arxiv.org/abs/2101.08176}{{\ttfamily 2101.08176}}.

\bibitem{Foreman:2021ljl}
S.~Foreman, T.~Izubuchi, L.~Jin, X.-Y.~Jin, J.C.~Osborn and A.~Tomiya,
  \emph{{{HMC}} with {{Normalizing Flows}}},
  \href{https://arxiv.org/abs/2112.01586}{{\ttfamily 2112.01586}}.

\bibitem{Foreman:2021ixr}
S.~Foreman, X.-Y.~Jin and J.C.~Osborn, \emph{Deep {{Learning Hamiltonian Monte
  Carlo}}},  \href{https://arxiv.org/abs/2105.03418}{{\ttfamily 2105.03418}}.

\bibitem{van1995python}
G.~Van~Rossum and F.L.~Drake~Jr, \emph{Python Tutorial}, {Centrum voor Wiskunde
  en Informatica Amsterdam, The Netherlands} (1995).

\bibitem{matplotlib}
T.A.~{Caswell}, M.~{Droettboom}, J.~{Hunter}, E.~{Firing}, A.~{Lee},
  J.~{Klymak} et~al., \emph{{matplotlib/matplotlib v3.1.0}},  May, 2019.
\newblock 10.5281/zenodo.2893252.

\bibitem{harris2020array}
C.R.~Harris, K.J.~Millman, S.J.~van~der Walt, R.~Gommers, P.~Virtanen,
  D.~Cournapeau et~al., \emph{Array programming with {NumPy}},
  \href{https://doi.org/10.1038/s41586-020-2649-2}{\emph{Nature} {\bfseries
  585} (2020) 357}.

\bibitem{tensorflow2015-whitepaper}
{Martín Abadi}, {Ashish Agarwal}, {Paul Barham}, {Eugene Brevdo}, {Zhifeng
  Chen}, {Craig Citro} et~al., ``{{TensorFlow}}: Large-{{Scale Machine
  Learning}} on {{Heterogeneous Systems}}.''

\bibitem{ipython4160251}
F.~{Perez} and B.E.~{Granger}, \emph{{IPython: A System for Interactive
  Scientific Computing}},
  \href{https://doi.org/10.1109/MCSE.2007.53}{\emph{Computing in Science and
  Engineering} {\bfseries 9} (2007) 21}.

\end{thebibliography}\endgroup

\end{document}